\documentclass[12p]{iopart}

\usepackage{graphicx}
\usepackage{color}
\usepackage{hyperref}
\usepackage{appendix}

\begin{document}

\title{Proton-neutron entanglement in the nuclear shell model} 

\author{Calvin W. Johnson}
\ead{cjohnson@sdsu.edu}
\author{Oliver C. Gorton}

\address{San Diego State University,
5500 Campanile Drive, San Diego, CA 92182-1233}


\begin{abstract}
We compute the 
proton-neutron entanglement entropy in the interacting nuclear shell model for a variety of nuclides and 
interactions. 
Some {results} make intuitive sense, for example that the shell structure, as governed 
by  single-particle and monopole energies, strongly affects the energetically available space and thus the entanglement entropy.  
We also find a surprising result: that the entanglement entropy at low excitation energy tends to decrease for nuclides when
$N \neq Z$. While we {provide evidence} this arises from the 
physical nuclear force {by contrasting with random two-body interactions which shows no 
such decrease}, the exact mechanism is  
unclear. Nonetheless, the low entanglement suggests that in models of neutron-rich nuclides, the 
coupling between protons and neutrons may be less computationally demanding than one might otherwise expect.
\end{abstract}


\submitto{\jpg}
\maketitle

\section{Introduction}

The structure of atomic nuclei exhibits a mixture of 
simple and complex behaviors.  What is meant by `simple' can be subtle, but typically it means the behavior
can be described by far fewer degrees of freedom than 
that required by modeling the nucleus as a collection of 
$A$ interacting nucleons; examples of simplicity include  algebraic 
models~\cite{talmi1993simple} and mean-field pictures~\cite{ring2004nuclear}. 
Of course, one must acknowledge that models themselves are not physical 
observables. Furthermore, complex models can mimic simpler ones, for example  quasidynamic symmetries~\cite{PhysRevC.58.1539,rowe2000quasi,bahri20003}, where a Hamiltonian mixes symmetries yet observables 
such as spectra and ratios of transition strengths are consistent with `simpler' 
symmetry-respecting models.

{Entanglement is a concept describing whether the observable coordinates of a quantum system are independent; whether measurement of one generalized coordinate $q_1$ influences future measurements of another coordinate $q_2$ of a system $\psi(q_1,q_2,...)$\cite{RevModPhys.80.517,Schroeder17}. Such correlations can be described by the entanglement entropy, a concept which has become popular in recent years due to increasing interest in quantum information and the potential of quantum computing~\cite{RevModPhys.81.865,strauch2016resource}. It is trivial to write down states which are either separable (not entangled) or in a superposition of separable states (entangled), but the creation of entangled states in nature relies on the existence of an interaction that mixes the relevant degrees of freedom.}


Here we consider the entanglement between the proton 
components and neutron components of configuration-interaction models of nuclei. Other recent work in entanglement entropy in nuclei addressed 
single-particle and seniority-mode entanglement~\cite{PhysRevC.103.034325,PhysRevC.106.024303} as well as orbital entanglement revealing shell closures~\cite{tichai2022combining}; we note the first two 
papers reference unpublished versions of the research reported here.

Although it does not directly correspond to the 
work here, we point out previous 
analyses of nuclear configuration-interaction wave functions using `entropy,'  
such as the 
configuration information entropy~\cite{PhysRevLett.74.5194,zelevinsky1995information}, which is simple but basis dependent, and the invariant correlation entropy~\cite{volya2003invariant}, 
which is much more complicated to compute. By contrast, because of the way our configuration interaction 
code constructs the wave functions, extraction of 
the wave function amplitudes in terms of proton-neutron 
coefficients is straightforward, a significant motivation for our approach.

In section \ref{SM} we lay out the basic framework of 
shell-model configuration-interaction calculations. 
In section \ref{ee}, we define  entanglement entropy as well as related concepts. 
We then provide examples of entanglement entropies for a variety of cases and show how much of the behavior for ground state entropies can be understood through standard concepts in nuclear structure physics.  A persistent phenomenon, however, is not so easily explained: realistic ground states of nuclides with $N \neq Z$ tend to have significantly smaller 
entanglement entropies than those with $N=Z$.  We also show 
trends for entropies for all states. 
We can show this is related to some components of 
realistic nuclear forces by contrasting them with results using  random interactions. While the mechanism for 
suppressing the entanglement eludes us, it is nonetheless worth reporting, not only as an apparently robust yet unexplained phenomenon, but also because it has a practical consequence: the low-lying states of 
neutron-rich nuclides have fewer nontrivial correlations between the proton and neutron components. 
This, in turn, suggests a practical approach for such nuclides, one which we are currently developing.

\section{The nuclear configuration-interaction shell model}

\label{SM}

{We find low-lying states of 
a nuclear Hamiltonian by the configuration-interaction method in a shell-model basis~\cite{BG77,br88,ca05}.}
Any many-body Hamiltonian can be written in second quantization  formalism as a polynomial in creation and annihilation operators~\cite{ring2004nuclear}:
\begin{equation}
        \hat{H} = \sum_{i}\epsilon_i \hat{a}_i^\dagger \hat{a}_i + \frac{1}{4} \sum_{ijkl}V_{ijkl}\hat{a}_i^\dagger \hat{a}_j^\dagger \hat{a}_l \hat{a}_k,
        \label{2ndquant}
\end{equation}
where $\epsilon_i$ are single particle energies and $V_{ijkl}$ are the two-body interaction matrix elements. {The single-particle operators $\hat{a}^\dagger_i$ create spin-1/2 nucleons in simple harmonic oscillator states with quantum numbers: $n_i$ (radial quantum number), $l_i$ (orbital angular momentum), and $j_i$ (total angular momentum). Many-body states are constructed as antisymmetrized products of these single particle states.}

 {To make calculations tractable, we limit the number of single particle valence states. For example, 
several of our calculation assume a fixed 
$^{16}$O core and allows valence nucleons in the $1s_{1/2}$-$0d_{3/2}$-$0d_{5/2}$ orbits, 
colloquially known as the \textit{sd} shell;
we also work in the $pf$-shell ($^{40}$Ca core 
with valence orbits $1p_{1/2,3/2}$-$0f_{5/2,7/2}$) and the combined $sd$-$pf$ shells. Starting from a finite single-particle valence space yields} a finite 
many-body basis~\cite{ca05}: 
\begin{equation}
    | \Psi \rangle = \sum_\alpha c_\alpha | \alpha \rangle, \label{CI}
\end{equation} 
where we use the occupation representation 
of Slater determinants, that is, of the form $| \alpha \rangle = \prod_i \hat{a}_i^\dagger | 0 \rangle .$ In particular we work in the $M$-scheme, {which means} the total 
$J_z$ or $M$ of all basis states is fixed to the same value. {For our calculations here 
we construct all possible valence configurations with fixed $M$.} 
Furthermore, we factorize the basis into proton ($\pi$) and neutron ($\nu$) components, so that 
we can write:
\begin{equation}
| \alpha \rangle = | \mu_\pi \rangle \otimes | \sigma_\nu \rangle. \label{bipartbasis}
\end{equation} 
This enables 
calculation of the proton-neutron entanglement entropy.

{The parameters $\epsilon_i$ and 
$V_{ijkl}$ in Eq.~(\ref{2ndquant}) are input parameters of the Hamiltonian. For details 
see the reviews in ~\cite{BG77,br88,ca05}. 
For our calculations we used both high-quality 
empirical interactions fitted separately 
in each model space to experimental spectra, 
as well as schematic interactions known to 
capture many features of nuclear structure, 
and randomly generated parameters. Using 
Eq.~(\ref{2ndquant}) and the factorized basis 
(\ref{CI}) one can compute~\cite{ca05,BIGSTICK} the matrix elements of the Hamiltonian in the many-body basis, $H_{\alpha, \beta} = \langle \alpha | \hat{H} | \beta \rangle.$}
Then the time-independent Schr\"odinger equation becomes a simple 
matrix eigenvalue problem: $\mathbf{H} \vec{c} = E \vec{c}$.

\section{Entanglement entropy}

\label{ee}

The entanglement entropy is a fundamental tool in quantum information science~\cite{RevModPhys.80.517,RevModPhys.81.865,strauch2016resource}.
Here we briefly review  {the development found in those sources.}

For a pure quantum state $| \Psi \rangle$, the density operator is 
$\hat{\rho} = | \Psi \rangle \langle \Psi |$; in a basis $\{ | \alpha \rangle \}$, 
i.e., Eq.~(\ref{CI}), the density matrix elements are $\rho_{\alpha^\prime \alpha} = 
c_\alpha^\prime c_\alpha^*$. Because this is idempotent, {$\rho^2 = \rho$},
and thus has either 0 or 1 as eigenvalues, the  von Neumann entropy, $S=-\tr(\rho\log\rho)$ vanishes. 

Suppose we work in a bipartite 
Hilbert space, e.g. $\mathcal{H}  = \mathcal{H}_\pi \otimes \mathcal{H}_\nu$, 
with basis states such as (\ref{bipartbasis}); we can then write Eq.~(\ref{CI}) explicitly as
\begin{equation}
    | \Psi \rangle = \sum_{\mu,\sigma} c_{\mu, \sigma} | \mu_\pi \rangle \otimes | \sigma_\nu \rangle. \label{pnbasis}
\end{equation} 
An unentangled state is one where one can transform to a basis where 
the amplitudes are separable, that is $c_{\mu, \sigma} = a_\mu b_\sigma;$ in this case the state could be written as a simple product: 
$| \Psi\rangle = \left( \sum_\mu a_\mu | \mu_\pi \rangle \right ) 
\otimes \left( \sum_\sigma b_\sigma | \sigma_\nu \rangle \right ).  $
{States which do not satisfy this are entangled.}

In the basis (\ref{pnbasis}) the density matrix is 
$\rho_{\mu^\prime \sigma^\prime, \mu \sigma} = c_ {\mu^\prime \sigma^\prime} c_{\mu \sigma}^*$. {This density matrix 
is idempotent.}
To get the \textit{reduced} density matrix, one 
traces over one of the subspace indices: 
\begin{equation}
    \rho^\mathrm{red}_{\mu^\prime, \mu} = \sum_\sigma 
    c_ {\mu^\prime \sigma} c_{\mu \sigma}^*.
\end{equation}
If the state is unentangled, the reduced density matrix will also 
be idempotent. {For a general state}
the reduced density matrix need not be idempotent, {and its eigenvalues 
can be between 0 and 1.} Then the 
entanglement entropy~\cite{RevModPhys.80.517}  
\begin{equation}
S_\mathrm{entangled} = - \mathrm{tr} \, \rho^\mathrm{red} \ln \rho^\mathrm{red}
\end{equation}
can be nonzero. {Because unentangled states must have 
zero entanglement entropy, non-zero entropy is a measure of entanglement~\cite{RevModPhys.80.517}.}
The fact that the eigenvalues of $ \rho^\mathrm{red}$ are 
real and non-negative, and independent of which subspace index we trace over, 
is a result of the singular value decomposition theorem; this is also called the Schmidt decomposition, especially in quantum information 
science.
The maximum entropy possible is the natural logarithm of the smaller
 subspace dimensions, 
 \begin{equation}
 S_{max} = \ln (\min ( \mathrm{dim}_\pi, \mathrm{dim}_\nu)).
 \end{equation}
Because our code is written using  an explicit proton-neutron basis, it is easy to extract 
$c_{\mu, \sigma}$ for any calculated state and then compute the entanglement 
entropy.

\section{Results}

 We work in three different model spaces and with several different shell model 
interactions. All of our calculations are in the $M$-scheme, that is, 
a basis with fixed total $J_z$. 
We start with several studies of $N=Z$ nuclides 
in the $sd$-shell, where our results can be 
easily understood. We then look at cases in the $sd$, $pf$, and $sd$-$pf$ spaces with
$N \neq Z$, which leads to a  surprise: 
ground states with $N \neq Z$ have significantly 
lower entanglement entropies than those with $N=Z$. By comparing with 
randomly generated interactions we provide evidence 
that this phenomenon has its origin in physical 
forces; but beyond that, we have yet to understand 
the specific mechanism.  Finally, we look at 
the behavior of the entropy over the entire spectrum.

\begin{figure}[ht]
\includegraphics[scale=0.45,clip]{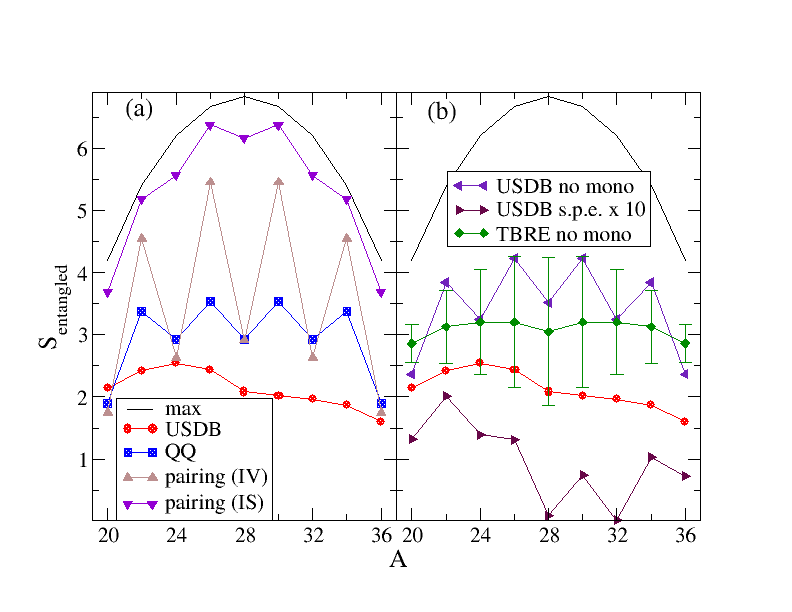}

\caption{Ground state entanglement entropy for $N=Z$ nuclides in the 
$sd$-shell.  Both panels show results for USDB~\cite{PhysRevC.74.034315}, a high-quality empirical interaction, and the maximum possible entanglement entropy. {In panel (a) } we show results  for 
an attractive isoscalar quadrupole-quadrupole (`QQ') interaction; isovector (IV) and isoscalar (IS) 
pairing. {In panel (b) we show results for}  USDB with single-particle energies and monopole interactions which have been set to zero, 
eliminating shell structure (`no mono'), and with  single-particle energies were inflated by a factor of ten, amplifying 
shell structure (`s.p.e. x 10'){. Additionally, panel (b)} gives the average and standard deviation for calculations drawn 
from a two-body random ensemble~\cite{PhysRevLett.80.2749}, also with shell structure eliminated (`TBRE no mono').
See text for discussion.}
\label{sdNeqZ}
\end{figure}

\begin{table}[]
    \centering
    \begin{tabular}{|c|c|}
    \hline
    Nuclide     &  dimension  \\
    \hline
     $^{20}$Ne, $^{36}$Ar    &  640 \\
     $^{22}$Na, $^{34}$Cl    &  6,116 \\
     $^{24}$Mg, $^{32}$S &   28,503 \\
          $^{26}$Al, $^{30}$P &   69,784 \\
     $^{28}$Si   &  93,710 \\
     \hline
    \end{tabular}
    \caption{$M=0$ dimensions for the $N=Z$ nuclides 
     in Fig.~\ref{sdNeqZ} with valence nucleons in the
    $sd$-shell.  Because there are a maximum of 12 valence nucleons of either species, the dimensions 
    have a particle-hole symmetry around N,Z=14.}
    \label{tab:dimensions}
\end{table}

\subsection{Examples with $N=Z$
and the role of the shell structure}

In this subsection we discuss some 
introductory results, focusing on 
$N=Z$ nuclides.  

\begin{figure}[ht]
\begin{center}
\includegraphics[scale=0.45,clip]{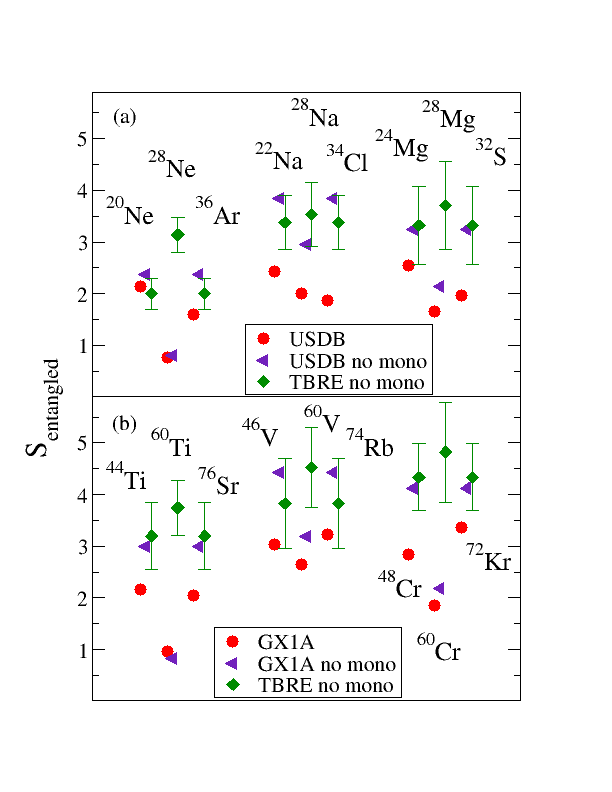}
\end{center}
\caption{Ground state entanglement beyond $N=Z$. Here we compare nuclide triplets 
with the same dimensionalities (and thus the same maximum entanglement entropy), with the same 
number of protons/proton holes and neutrons/neutron holes; not all cases correspond to physical nuclides. Panel (a) is for $sd$-shell nuclides, while panel (b) is for $pf$-shell nuclides. Here USDB~\cite{PhysRevC.74.034315} is an empirical 
interaction for the $sd$-shell while GX1A~\cite{PhysRevC.65.061301,PhysRevC.69.034335,honma2005shell} is for the $pf$-shell. We also show the average and 
standard deviation for  interactions drawn from the two-body random ensemble (TBRE). Finally, `no mono' means the single-particle energies 
and monopole interaction terms have been set to zero, thus eliminating any shell structure~\cite{ca05}.}
\label{fig:conj}
\end{figure}

Our first examples in Fig.~\ref{sdNeqZ} are
the ground state entanglement entropies for $N=Z$ nuclides in the so-called 
$sd$-shell, which has a fixed $^{16}$O core and valence particles in the 
$1s_{1/2}$-$0d_{3/2}$-$0d_{5/2}$ orbitals. {For convenience we give the $M=0$ dimensions in Table~\ref{tab:dimensions}.} Here the proton and neutron 
many-body spaces have equal dimensions. We use a high-quality empirical interaction, the universal $sd$-shell interaction, version B or USDB, {which, like all similar empirical interactions, is represented as a list of single-particle energies and two-body matrix elements fitted to data}~\cite{PhysRevC.74.034315}.
We also show in Fig.~\ref{sdNeqZ}(a) the entropies for ground states of the attractive isoscalar quadrupole-quadrupole (QQ) interaction, 
the attractive isovector (IV) pairing, 
that is, nucleons paired up to isospin $T=1$, and the attractive 
isoscalar (IS) pairing,
or nucleons paired up to $T=0$. 
{
These schematic interactions are well-known in nuclear structure 
physics~\cite{BG77,ring2004nuclear,bohr1998nuclear1}; we give their 
exact definitions in the Appendix.}
Unsurprisingly, 
isoscalar pairing, which forces protons to pair with neutrons, has nearly maximal 
entanglement entropy. Isovector pairing shows a strong odd-even effect: odd-odd 
nuclides, where at least one proton and at least one neutron must pair up, has 
much higher entropy than the even-even cases. Ground states of the QQ interaction 
have a much weaker odd-even staggering, while USDB ground states have the lowest 
entropies of all and exhibit no odd-even staggering.

In Fig.~\ref{sdNeqZ}(b), we further investigate the origin of some of these 
behaviors. The shell structure of nuclei is governed by the single-particle 
energies and the so-called monopole terms~\cite{ca05}, that is, terms in 
the interaction of the form $\hat{n}_a \hat{n}_b$ where $\hat{n}_a$ is the number 
operator for orbital $a$. By setting the single-particle energies and monopole 
terms to zero, the ground state entanglement entropy increases, and shows an 
odd-even staggering comparable to QQ.  Conversely, by inflating the {standard USDB values of the} single-particle energies
{($\epsilon(1s_{1/2}) = -3.2079$ MeV, $\epsilon(0d_{3/2}) =  2.1117$ MeV, and $\epsilon(0d_{5/2}) =   -3.9257$ MeV)}
by a factor of $\times 10$, we restrict the space energetically 
available and dramatically decrease the ground state entanglement entropy, 
reaching zero at shell closures.  Finally, we considered 
two-body interactions drawn from a random ensemble (TBRE~\cite{PhysRevLett.80.2749}), 
also removing single-particle energies and monopole terms. We used results from ten 
different members of the ensemble to get average entropies and corresponding standard deviations. 
From this we learn that, first, the shell structure, as encoded in single-particle 
energies and monopole terms, has a strong effect {on the entanglement entropy}, by energetically restricting 
the available model space {and thus reducing the 
effective dimension}. While the `no mono' USDB results appear 
consistent 
 with a randomly drawn interaction, it does show a nontrivial odd-even staggering.  Later, we 
will see additional behaviors that strongly differ between USDB and randomly 
generated interactions.

{From these numerical experiments we 
conclude the lower entanglement of the full USDB wave functions is due to the shell structure, i.e., the monopole terms and single-particle energies. We speculate the lack of odd-even staggering in the entropies of full USDB calculations may be due to many small, quasi-random components in the interaction beyond pairing and QQ.}

\subsection{Away from $N=Z$}

In the previous subsection we considered exclusively 
nuclides with $N=Z$. Here we compare entropies 
for nuclides with $N \neq Z$, and find 
systematically lower entropies--albeit when using 
physical forces. {Because we use interactions which respect isospin as a good 
symmetry, we only consider here $N > Z$.}

\begin{figure}[ht]
\includegraphics[scale=0.45,clip]{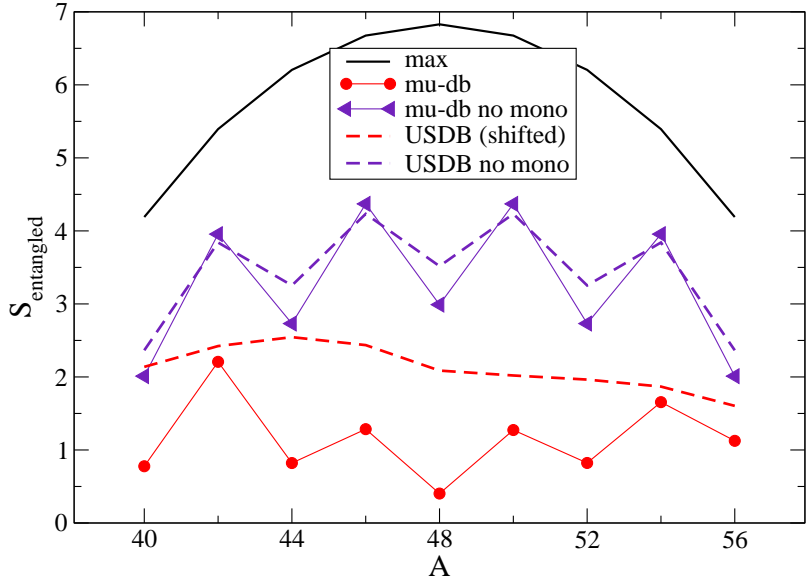}

\caption{Entanglement entropies for nuclides in the $sd$-$pf$ space, from $^{40}$Ne through $^{56}$Ar. See 
text for a detailed description of the model space, which is truncated so that the dimensionalities 
are the same as for $N=Z$ $sd$-shell nuclides in Fig.~\ref{sdNeqZ}. Here `mu-db' is the interaction of 
Ref.~\cite{PhysRevLett.116.112502}, while `max' and `no mono' mean the same as in Fig.~\ref{sdNeqZ}. We also include  for comparison  
$sd$-shell results from Fig.~\ref{sdNeqZ} (dashed lines), shifted by $A=20$} 
\label{sdpf}
\end{figure}

In Fig.~\ref{fig:conj} we compare the ground 
state entropies for `triplets' of nuclides. 
These triplets are set in the same valence space, 
either $sd$ or $pf$, and all members of a triplet have the same number of 
valence protons or proton holes, and the same number of 
valence neutrons or neutron holes. For example, we have 
$^{20}$Ne, with 2 valence protons and 2 valence neutrons; $^{36}$Ar, with 2 proton holes and 2 neutron holes (since the filled $sd$ valence orbitals contain 12 particles of a given nuclear species); and $^{28}$Ne, with 2 valence protons and 2 neutron holes. By this construction all members of a triplet have exactly the same dimensionalities.  For the $sd$-shell cases in Fig.~\ref{fig:conj}(a), following 
Fig.~\ref{sdNeqZ}, we compare entanglement entropies from ground states computed 
with the full USDB interaction, USDB with monopole terms and single-particle 
energies set to zero, and finally entropies of ground states calculated 
from the two-body random ensemble (with monopole terms zeroed out). 
For the $pf$-shell cases in Fig.~\ref{fig:conj}(b), we do the same but 
replace USDB with the GX1A interaction~\cite{honma2005shell}. 

We see a strong and persistent trend: ground states of 
nuclides with $N \neq Z$ computed with realistic interactions 
have significantly lower entanglement than those with $N=Z$, a result 
we have replicated in other shell model spaces we do not show. 
To put this in perspective, an entropy difference of 1 corresponds to a 
difference in effective dimensionality of $e=2.71\ldots$. 
This trend 
is stronger for even-even nuclides and when the shell structure is removed. 
For ground states computed under the TBRE, however, that trend disappears 
and is even slightly reversed. 
{While we have made attempts to devise a 
plausible model for these behaviors, for example why the $N\neq Z$ entropy is lower for realistic interactions but is higher for the TBRE, we have not 
succeeded.}

Because this trend suggests a lower entanglement for neutron-rich nuclides (or, because of isospin symmetry for these interactions, proton-rich as well), 
we continue our investigation in 
 Fig.~\ref{sdpf}, where we consider cross-shell examples in
 the $sd$-$pf$ space using the `mu-db' interaction~~\cite{PhysRevLett.116.112502}.
 In order to compare with past results, we restrict the space so as to 
 follow the dimensionalities of Fig.~\ref{sdNeqZ}. Thus 
 we restrict protons 
to the $sd$-shell, while  for neutrons the 
$sd$-shell and the $0f_{7/2}$ orbitals are filled 
and frozen, leaving only $0f_{5/2}$-$1p_{3/2}$-$1p_{1/2}$ as the active 
space for neutrons. Finally, we restrict ourselves to 
the same number of active protons and neutrons.
Thus the nuclides correspond to $^{40}$Ne, 
$^{42}$Na, $^{44}$Mg, and so on, through $^{56}$Ar. These 
restrictions are chosen so that the dimensionalities are the same 
as in Fig.~\ref{sdNeqZ}.  
While somewhat artificial--we do not claim all 
these correspond to physical nuclides--this 
nonetheless allows us to make a clean investigation 
into the entanglement. 
For ease of comparison, 
we include the USDB and USDB monopole-subtracted (`no mono') 
results for $N=Z$ nuclides, shifted over by $A=20$ (so that $^{20,40}$Ne are superimposed, 
etc.).
Again we see a low entanglement entropy, even lower than for 
the $sd$-shell examples in Fig.~\ref{sdNeqZ}, although the `no monopole' case shows much of 
this is driven by shell structure.  Nonetheless 
this provides evidence that the low-entanglement for 
$N \neq Z$ nuclides persists for cross-shell spaces.

It is important to note that this is not simply 
isospin. Although we do not show it, for a 
given $N$ and $Z$, states of different $J$ and $T$ 
nonetheless show very similar trends. 
In other words, this behavior is related to 
$T_z$, not $T$.   In the next section we investigate 
further.

\subsection{Entanglement entropies of excited levels}

In the previous results we focused on ground 
state entanglement.  Here we look at systems where we 
can fully diagonalize and compute the entanglement entropy 
across the spectrum.

\begin{figure}[ht]
\includegraphics[scale=0.45,clip]{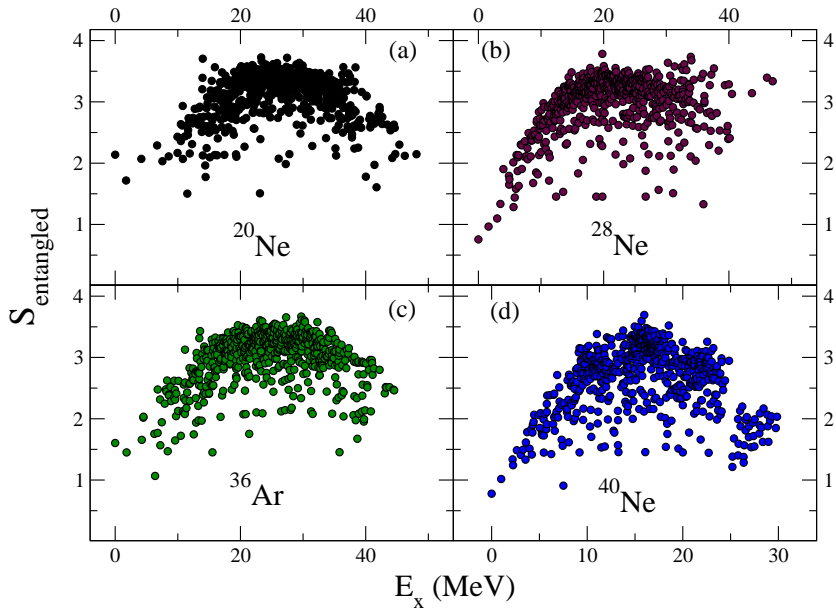}

\caption{Entanglement entropies for all levels as a function of excitation energy {$E_x$}. 
Panels (a), (b), and (c) are computed in the $sd$-shell using the USDB interaction~\cite{PhysRevC.74.034315}.  Panel (d) is computed in the $sd$-$pf$ space, 
using the `mu-db' interaction~\cite{PhysRevLett.116.112502}; certain orbits are frozen in order to make the 
dimensions for this case the same as the other three cases.
The maximum entanglement entropy is at the top of each plot. \label{sdspectrum}}
\end{figure}

In Fig.~\ref{sdspectrum} we present the entanglement entropy for all levels for 
several nuclides computed with empirical interactions that give a good description of the data. 
Specifically we consider $^{20,28}$Ne and $^{36}$Ar, computed in the $sd$-shell with the USDB 
interaction~\cite{PhysRevC.74.034315},
and $^{40}$Ne computed in a truncated $sd$-$pf$ space, computed with 
a cross-shell interaction~\cite{PhysRevLett.116.112502}; for the latter, we restricted the valence 
protons to the $sd$-shell, and froze all neutrons in the $sd$-shell and in the $0f_{7/2}$ orbital, 
so that we have only two valence neutrons in the $0f_{5/2}$-$1p_{3/2}$-$1p_{1/2}$ orbitals. 
These choices are made so that all four cases have exactly the same dimensionalities. 
While each plot has considerable scatter, there are a couple of noticeable features. 
The first is an overall curvature: on average, the entropy rises and then falls. 
The second feature is that the isospin-asymmetric nuclides, $^{28,40}$Ne, are noticeably more 
asymmetric in the distribution of entropies, in particular the low entropy of the ground state 
as pointed out in the previous section.

\begin{figure}[ht]
\includegraphics[scale=0.45,clip]{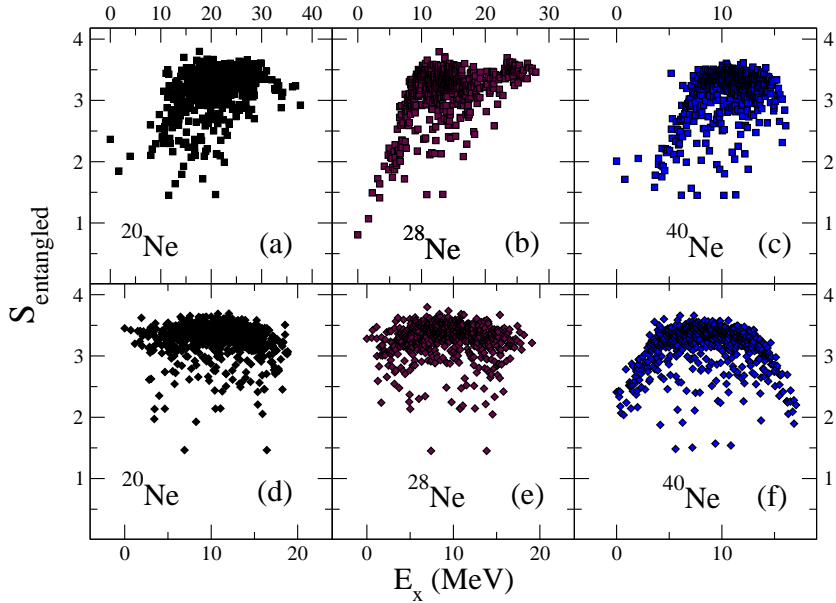}

\caption{Entanglement entropies for all levels as a function of excitation energy {$E_x$}. 
The top panels, (a)-(c), are computed with realistic interactions (USDB for panels (a) and (b), and `mu-db' for panel (c)) but with all single-particle 
energies and monopole interaction terms set to zero, so as to remove shell effects. The 
bottom panels, (d)-(f), correspond to the same nuclides, but with a random two-body 
interaction. 
The maximum entanglement entropy is the top of each plot.}
\label{trless}
\end{figure}

To probe the origin of these behaviors, in Fig.~\ref{trless} we recomputed the entropies 
for $^{20,28,40}$Ne. In the top panels, Fig.~\ref{trless}(a)-(c), we set the single-particle 
energies and monopole interaction terms to zero. (Note that in such a scenario, there is a particle-hole symmetry, so that $^{36}$Ar would be exactly the same as $^{20}$Ne.)  
This has the effect of amplifying the asymmetry in all three cases.  In the bottom panels, 
Fig.~\ref{trless}(d)-(f), we instead generated a random set of two-body matrix elements, and also 
removed the monopole interaction terms. 
Here the asymmetry vanishes, and the curvature for $^{20,28}$Ne is much reduced. An important 
lesson we learn is that the $T_z$ dependence of the entropy appears to come out of the physical 
nuclear force, as it does not appear when using a randomly generated interaction.

\section{Conclusions and acknowledgements}

We have found that both schematic and realistic nuclear shell model Hamiltonians 
have low entanglement between the proton and neutron components of the wave function. 
This is particularly pronounced for states with high isospin.  Part of the behavior is 
governed by the shell structure, which reduces the effective dimensionality, but, by 
using  random two-body matrix elements, we can establish that the low entropy also is a 
feature of typical nuclear forces. While we have spent considerable 
effort to construct toy models to understand this behavior,  
so far none of them have provided convincing illumination.

A low entanglement means that one can get a good approximation to a wave function 
using a much smaller subset of basis states. This is the driving idea between 
the density matrix renormalization group methods~\cite{PhysRevLett.69.2863,PhysRevB.48.10345,RevModPhys.77.259}, which 
have only been used sporadically in nuclear structure physics~\cite{tichai2022combining,PhysRevC.65.054319,PhysRevLett.97.110603,PhysRevC.78.041303,PhysRevC.92.051303}, partitioning 
on orbitals rather than between protons and neutrons. Closer to the present work is
the proton-neutron singular-value 
decomposition analysis of shell model wave functions~\cite{PhysRevC.67.051303,PhysRevC.69.024312}; ironically, the latter studies focused on $N \approx Z$ nuclides.
One happy conclusion from our work presented here is that reduced basis methods, justified by 
low entanglement~\cite{PhysRevE.84.056701}, may be even more 
effective for high-isospin nuclides, such as heavy nuclei, where the need for 
dimensional reduction is greatest.  We have made progress in a systematic implementation 
of this idea and will present results soon.

This material is based upon work supported by the U.S. Department of Energy, Office of Science, Office of Nuclear Physics, 
under Award Number  DE-FG02-03ER41272, and by the Office of High Energy Physics, under Award No.~DE-SC0019465, and by Lawrence Livermore National Laboratory under Contract DE-AC52-07NA27344, with support from the ACT-UP award.

\appendix

\section{Schematic interactions}
{
Schematic interactions are well-known to capture many features 
of nuclear structure and have long been used in nuclear physics~\cite{BG77,ring2004nuclear,bohr1998nuclear1}.  For completeness we summarize them here.

\medskip

\noindent $\bullet$  Quadrupole-Quadrupole (QQ): 
We construct a quadrupole operator
\begin{equation}
    \hat{Q}_{2M,TM_T}
    = \sum_{ab} Q_{ab} [ \hat{c}^\dagger_a \otimes \tilde{c}_b]_{2M, TM_T}
\end{equation}
where $a,b$ are indices for single-particle orbits 
defined by $n, l, j$; $[ \cdot \otimes \cdot ]_{JM,TM_T}$
indicates coupling by Clebsch-Gordan coefficients up to 
total angular momentum $J$ (here $J=2$) and $z$-component $M$ and total isospin $T$ and third component $M_T$;
$\hat{c}^\dagger_{a}$ creates a 
particle in orbit $a$, $\tilde{c}_b$ is a time-reversed 
destruction operators~\cite{edmonds1996angular} for a particle in orbit $b$; 
and finally the reduced matrix elements~\cite{edmonds1996angular} of the quadrupole 
operator (which we compute in a harmonic-oscillator basis)
\begin{equation}
Q_{ab} = \langle a || r^2 Y_2 || b \rangle,
\end{equation}
where $Y_{lm}(\theta,\phi)$ is a spherical harmonic.
 The quadrupole-quadrupole Hamiltonian is 
\begin{equation}
  V_{QQ}  \left [ \hat{Q} \otimes \hat{Q} \right ]_{00,00},
\end{equation}
where $V_{QQ}$ is the strength of the interaction and is 
taken $< 0$ to make it attractive; because we only focus on the wave functions and not the energies, the magnitude of $V_{QQ}$ is unimportant here. For our calculations we considered only the isoscalar (T=0) QQ interaction.

\medskip

\noindent $\bullet$  Pairing. The  
pairing operator is 
\begin{equation}
    \hat{P}^\dagger_{T M_T} = \sum_a \sqrt{2j_a+1} [\hat{c}^\dagger_a 
    \otimes \hat{c}^\dagger_a]_{00,T M_T},
\end{equation}
where $j_a$ is the angular momentum of orbital $a$.
The pairing Hamiltonian is 
\begin{equation}
    G \sum_{M_T} (-1)^{T-M_T} \hat{P}^\dagger_{T M_T} 
    \hat{P}_{T M_T},
\end{equation}
where $G$ is the strength of the pairing Hamiltonian,
and is taken $< 0$ to make attractive. Here $T=1$ yields
isovector pairing and $T=0$ isoscalar pairing.

}

\bigskip

\bibliographystyle{unsrt}
\bibliography{johnsonmaster}

\end{document}